\pdfoutput=1
\documentclass[11pt]{article}

\PassOptionsToPackage{hyphens}{url} 
\usepackage[preprint]{acl}
\usepackage{times}
\usepackage{latexsym}
\usepackage[T1]{fontenc}
\usepackage[utf8]{inputenc}
\usepackage{microtype}
\usepackage[varqu]{inconsolata}
\usepackage{graphicx}
\usepackage{amsmath,amssymb}
\usepackage{booktabs}
\usepackage{multirow}
\usepackage{makecell}
\usepackage{array}
\usepackage{xcolor}
\usepackage{colortbl}
\usepackage{listings}
\usepackage{tikz}
\usepackage[most]{tcolorbox}
\usepackage{tabularx}
\usepackage{ragged2e}
\usepackage{etoolbox}
\apptocmd{\thebibliography}{\RaggedRight}{}{\PackageWarning{main}{could not patch thebibliography}}
\usetikzlibrary{positioning,arrows.meta,fit,calc,backgrounds,shadows,decorations.pathreplacing}

\definecolor{snow}{HTML}{F7FAFC}
\definecolor{sky}{HTML}{A7C7E7}
\definecolor{ice}{HTML}{CFEAF6}
\definecolor{coral}{HTML}{F4A7AE}
\definecolor{peach}{HTML}{FBD9C9}
\definecolor{navy}{HTML}{5878A6}
\definecolor{mint}{HTML}{A6D0C6}
\definecolor{mist}{HTML}{DADDE3}
\colorlet{ink}{navy!55!black}
\colorlet{grp}{mist!35!ink}
\colorlet{cid}{mint!55!ink}\colorlet{crec}{coral!70!ink}\colorlet{cfmt}{navy}
\newcommand{\cn}[2]{\tikz[baseline=(c.base)]\node[circle, fill=#1, inner sep=0pt, minimum size=1.25em,
  font=\sffamily\bfseries\scriptsize, text=white] (c) {#2};}
\newcommand{\pill}[2]{\tikz[baseline=(p.base)]\node[fill=#1, rounded corners=2pt, inner xsep=2.4pt, inner ysep=1pt,
  font=\scriptsize, text=ink] (p) {#2};}
\newtcblisting{recipe}[1][]{enhanced, listing and comment, colback=snow, colframe=mist!80!ink, boxrule=0.5pt,
  arc=2.5pt, left=4pt, right=3pt, top=3pt, bottom=3pt, middle=3pt,
  segmentation style={draw=mist!80!ink, dashed, line width=0.4pt},
  colbacktitle=mist!55!white, coltitle=ink, fonttitle=\small, toptitle=1.5pt, bottomtitle=1.5pt, lefttitle=0pt,
  titlerule=0pt,
  title={\tikz[baseline=(t.base)]\node[fill=navy, text=white, font=\small\bfseries, inner xsep=5pt, inner ysep=2.5pt,
    rounded corners=1.5pt] (t) {Recipe};\enspace labels for fine-tuning the Qwen3-Omni Talker},
  listing options={language=Python, basicstyle=\ttfamily\footnotesize, columns=fixed, basewidth=0.5em,
    keywordstyle=\color{navy}\bfseries, commentstyle=\color{grp}, stringstyle=\color{coral!65!ink},
    emph={F, loaders, mimi}, emphstyle=\color{ink}, upquote=true, showstringspaces=false, keepspaces=true,
    escapeinside={(*@}{@*)}, aboveskip=0pt, belowskip=0pt, frame=none, backgroundcolor={}}, #1}
\newcommand{\loudtwo}[2]{\makebox[2.1em][r]{#1/#2}\,\tikz[baseline=-0.55ex]{%
  \fill[mist!70!white] (0,0.008) rectangle (0.5,0.06); \fill[coral!82!ink] (0,0.008) rectangle ({0.5*#1/60},0.06);
  \fill[mist!70!white] (0,-0.06) rectangle (0.5,-0.008); \fill[coral!82!ink] (0,-0.06) rectangle ({0.5*#2/60},-0.008);}}
\newcommand{\loudone}[1]{\makebox[2.1em][r]{#1}\,\tikz[baseline=-0.55ex]{%
  \fill[mist!70!white] (0,-0.06) rectangle (0.5,0.06); \fill[coral!82!ink] (0,-0.06) rectangle ({0.5*#1/60},0.06);}}

\newcommand{\agree}[1]{#1\,\tikz[baseline=-0.55ex]{\fill[mist!70!white] (0,-0.06) rectangle (0.62,0.06);
  \fill[navy!70] (0,-0.06) rectangle ({0.62*#1},0.06);}}
\newcommand{\ownbox}[1]{\tikz[baseline=(o.base)]\node[draw=coral!75!ink, line width=0.6pt, rounded corners=1.5pt,
  inner xsep=1.4pt, inner ysep=0.8pt] (o) {#1};}
\arrayrulecolor{ink}

\newcommand{\code}[1]{\texttt{#1}}
\title{Reproducing the Speech Labels of Qwen3-Omni's Talker:\\ Mimi in bf16, and a Last-Frame Pitfall}
\author{Yuxuan Chen \\
  The Chinese University of Hong Kong, Shenzhen \\
  \texttt{yuxuanchen5@link.cuhk.edu.cn}}

\begin{document}
\maketitle

\begin{abstract}
Qwen3-Omni speaks through a Talker predicting 16 codebooks of a 12.5\,Hz audio codec, but the encoder that made its
training labels was not released, so new voices cannot be trained on matching labels. We identify this encoder from
the released model's behavior. White-box regression shows that the codebook embeddings of the Talker, its multi-token
predictor and the Code2Wav decoder match the first 16 codebooks of kyutai's Mimi layer by layer. A decoder-based test
looks convincing but fails calibration: decoders trained on known labels cannot tell them apart. Teacher-forced Talker
likelihoods, calibrated with 13 code language models trained on known labels, do tell implementations apart: the
Talker prefers labels from kyutai's \code{moshi} package with bf16 weights over every other variant we tried. Loudness
normalization and band-limiting of 20\% or more of the data are ruled out. Separately, transformers' \code{MimiModel}
encodes the last frame differently from kyutai's encoder unless the input length is a multiple of 1920 samples. After
LoRA fine-tuning on its labels, 44--55\% of generated sentences end in a noise burst, and the same frame costs the
Mimi-based CSM-1B 119 nats; padding the input removes both. We release the encoding recipe, a label checker and a
Talker fine-tuning toolkit.
\end{abstract}

\section{Introduction}

\begin{figure*}[t]
\centering
\includegraphics{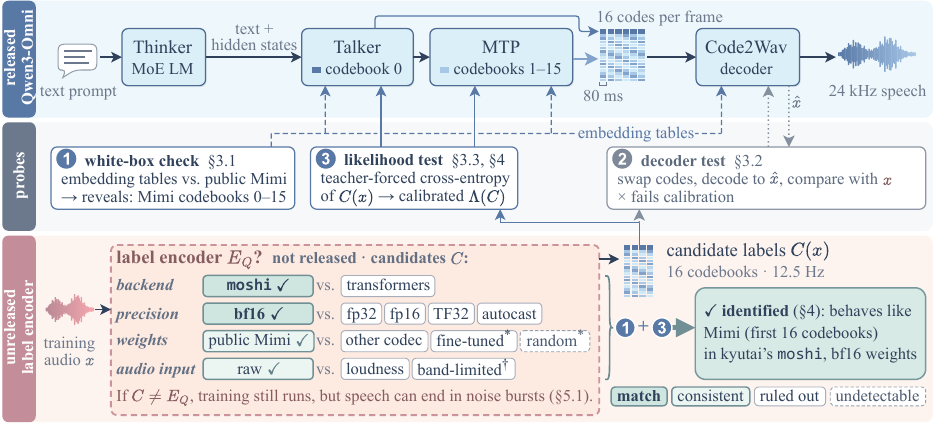}
\caption{Qwen3-Omni's released speech path (top), the unreleased label encoder $E_Q$ that we identify (bottom), and the
three probes that connect them (middle). The encoder that made the Talker's and Code2Wav's training labels is not
released; we test candidate encoders $C$ in its place, one setting per row (Sections~\ref{sec:setup}
and~\ref{sec:impl}). ``Consistent'' chips fit the evidence but are not pinned down. Probe~1, the white-box check
(Section~\ref{sec:whitebox}),
compares the codebook embedding tables of the Talker, the MTP and Code2Wav with the public Mimi codebooks and rules out
other codecs. Probe~2, the decoder test (Section~\ref{sec:decoder}), swaps single codes, decodes them with Code2Wav
into $\hat x$ and asks whether a candidate's quantization margins predict how far $\hat x$ moves from the audio $x$; it
fails calibration, so we do not use it as evidence (grey). Probe~3, the likelihood test (Sections~\ref{sec:likelihood}
and~\ref{sec:impl}), scores $C(x)$ under the Talker (codebook 0) and the MTP (codebooks 1--15) with teacher forcing and
rules out the other plain chips. $^{*}$Only our two fine-tuning directions are ruled out; random weight perturbations
that change 2--10\% of the codes in codebooks 1--4 are not detected, so we identify behavior, not bit-exact weights.
$^{\dagger}$Only band-limiting of 20\% or more of the data is ruled out. In the code grids, the dark top row is
codebook 0 and the other rows are codebooks 1--15.}
\label{fig:pipeline}
\end{figure*}

Qwen3-Omni~\citep{qwen3omni} generates speech in two stages (Figure~\ref{fig:pipeline}). The Thinker, a
mixture-of-experts language model, writes the text; the Talker, a smaller mixture-of-experts model, turns the Thinker's
text and hidden states into audio codes, one frame every 80\,ms. Each frame has 16 codebooks: the Talker predicts the
first, a multi-token predictor (MTP) predicts the other 15, and a convolutional decoder (Code2Wav) turns codes into a
24\,kHz waveform. The released checkpoint contains all three generators but not the encoder that turned training audio
into codes. Without it, nobody outside Qwen can make labels to fine-tune the Talker, which is the part that decides
what the model sounds like.

The question has come up repeatedly in public without an official answer; the community found that Qwen3-Omni codes
decode with the decoder of kyutai's Mimi~\citep{moshi} and already uses Mimi to make labels
(Section~\ref{sec:related}).

``It decodes with Mimi'' leaves open how close the labels are to Mimi's encoder as one would run it today: same
weights or fine-tuned ones, fp32 or a lower precision, the reference \code{moshi} code or the transformers port, raw or
preprocessed audio. On codebooks 1--4, bf16 inference changes 12\% of the codes, \code{moshi} instead of transformers
(both in bf16) about 15\%, band-limiting to 16\,kHz 40\%, and loudness normalization 67\%. We answer these questions
with tests whose sensitivity we first measure on models trained with known labels. Our contributions:

\begin{enumerate}
  \item Identification of the label encoder: white-box evidence that the Talker's codebooks are Mimi's first 16,
        and a calibrated likelihood test that singles out Mimi run with kyutai's \code{moshi} package and bf16 weights.
  \item A negative result: a decoder-based test that ranks the public encoder first is uninformative, because
        decoders trained on known labels show the same ranking.
  \item A last-frame mismatch between transformers' Mimi encoder and kyutai's that makes about half of the
        sentences from fine-tuned Qwen3-Omni and CSM-1B end in a noise burst.
  \item An encoding recipe that pads the input to avoid it, a label checker and a Talker fine-tuning toolkit.
\end{enumerate}

\begin{figure}[t]
\begin{recipe}[comment={\footnotesize\renewcommand{\arraystretch}{1.18}%
\begin{tabularx}{\linewidth}{@{}c@{\hspace{4pt}}X@{\hspace{3pt}}r@{}}
\cn{cid}{1} & kyutai's \code{moshi} loader & \pill{mint!45!white}{identified \S\ref*{sec:impl}} \\
\cn{cid}{2} & weights in bf16 & \pill{mint!45!white}{identified \S\ref*{sec:impl}} \\
\cn{crec}{3} & zero-pad the tail & \pill{peach!80!white}{recommended \S\ref*{sec:lastframe}} \\
\cn{cfmt}{4} & 16 codebooks, $\lceil T/1920\rceil$ frames & \pill{ice}{format \S\ref*{sec:whitebox}} \\
\end{tabularx}\\[1pt]
{\scriptsize\color{grp}With transformers' \code{MimiModel}, pad the input to a multiple of 1920 samples.}}]
import torch
import torch.nn.functional as F
from moshi.models import loaders

mimi = loaders.get_mimi(weight_path,    (*@\cn{cid}{1}@*)
                        device="cuda",
                        num_codebooks=16)
mimi = mimi.to(torch.bfloat16).eval()   (*@\cn{cid}{2}@*)

@torch.no_grad()
def encode(wav):  # [batch, 1, T], 24 kHz
    n = -(-wav.shape[-1] // 1920)       (*@\cn{cfmt}{4}@*)
    x = F.pad(wav, (0, wav.shape[-1]))  (*@\cn{crec}{3}@*)
    x = x.to(torch.bfloat16)
    return mimi.encode(x)[:, :16, :n]   (*@\cn{cfmt}{4}@*)
\end{recipe}
\caption{Recipe for labels that match the Qwen3-Omni Talker. Green callouts mark what we identified in the released
model, the coral one our recommendation, the blue ones the label format.}
\label{fig:recipe}
\end{figure}

Figure~\ref{fig:recipe} gives the resulting recipe.\footnote{\RaggedRight Code: \url{https://github.com/elpsykongloo/qwen3-omni-talker-finetune} (recipe,
\code{label\_check}, \code{talker\_tune}). Data:
\url{https://huggingface.co/collections/mifanbushipeicai/qwen3-omni-talker-finetune-6aec41e50ac465e3f0ef9222}
(test-set codes for all variants, Talker log-likelihoods, the 13 calibration models, sample voices).} Its parts have different status: the encoder and its precision are
what we identified in the released model; the tail padding is our recommendation, since the last frame says nothing
about how Qwen encoded (Section~\ref{sec:impl}). It matters with transformers' \code{MimiModel}, which otherwise
zero-pads hidden states inside the encoder; the resulting last-frame codes often decode as a noise burst that
fine-tuned models learn to reproduce (Section~\ref{sec:lastframe}). We padded each utterance to twice its length.
Padding a batch to its longest item leaves that item unpadded: with batches of 8, 6\% of utterances still ended in a
burst instead of 1\%. For LoRA voice adaptation the precision did not matter in our tests,
the padding did.

\section{Setup}\label{sec:setup}

\paragraph{Notation and hypotheses.} $E_Q$ is the encoder Qwen used to make the Talker's labels (not the Thinker's
audio encoder). Our reference is the public Mimi encoder (\code{kyutai/mimi}, first 16 of its 32 codebooks) run in
fp32 with transformers~\citep{transformers}; we call its output the \emph{fp32 reference codes}. We test four
hypotheses: (i)~\emph{same encoder}: $E_Q$ is the public encoder, possibly run with a different numerical
implementation (precision, library); (ii)~\emph{same encoder with preprocessing}: (i) plus a fixed preprocessing step
such as band-limiting or loudness normalization; (iii)~\emph{modified encoder}: the same quantizer and code ids, but a
fine-tuned or replaced encoder network; (iv)~\emph{different tokenizer}: an unrelated tokenizer with the same tensor
shapes.

\paragraph{Candidate encoders.} Besides the reference we construct encoders that are close to it in known ways
(Table~\ref{tab:lambda}, first column): \emph{random perturbations} of the weights that change 2\%, 5\% or 10\% of
codes in codebooks 1--4 (named by that target, e.g.\ ``random 5\%''), \emph{fine-tuning points} that move the weights
along a real fine-tuning direction on Chinese or English speech, the transformers model in bf16, band-limited input
and loudness-normalized input. Section~\ref{sec:impl} adds implementation variants. Appendix~\ref{app:candidates}
describes the construction and Appendix~\ref{app:variants} lists how every variant is produced. Candidates were built
without looking at any Qwen score.

\begin{table*}[t]
\centering\small
\setlength{\tabcolsep}{3pt}
\renewcommand{\arraystretch}{1.12}
\begin{tabular}{@{}lll rrrrr @{\hspace{9pt}} r@{}}
\toprule
 & & & \multicolumn{5}{c}{$\Lambda$ under a calibration model trained on labels from} & \\
\cmidrule(lr){4-8}
\multicolumn{2}{@{}l}{candidate encoder} & agreement & \thead{fp32\\{\scriptsize seed 1 {\color{grp}/ seed 2}}} &
\thead{random\\5\%} & \thead{transformers\\bf16} & \thead{band-\\limited} & \thead{band-limited,\\bf16} &
\thead{\textbf{Talker}\\$\Lambda$} \\
\midrule
\multirow{3}{*}{{\color{grp}\makecell[l]{random\\perturbation}}} & 2\% & \agree{0.979} & \cellcolor{sky!16!white}$+0.051$\,{\scriptsize\color{grp}/\,$0.052$} & \cellcolor{coral!33!white}$\mathbf{-0.096}$ & \cellcolor{sky!13!white}$+0.038$ & \cellcolor{sky!24!white}$+0.114$ & \cellcolor{sky!19!white}$+0.070$ & \cellcolor{sky!9!white}$+0.026$ \\
 & 5\% & \agree{0.948} & \cellcolor{sky!22!white}$+0.096$\,{\scriptsize\color{grp}/\,$0.095$} & \cellcolor{coral!45!white}\ownbox{$\mathbf{-0.270}$} & \cellcolor{sky!20!white}$+0.074$ & \cellcolor{sky!22!white}$+0.091$ & \cellcolor{sky!23!white}$+0.099$ & \cellcolor{sky!7!white}$+0.023$ \\
 & 10\% & \agree{0.905} & \cellcolor{sky!32!white}$+0.239$\,{\scriptsize\color{grp}/\,$0.241$} & \cellcolor{coral!14!white}$-0.007$ & \cellcolor{sky!25!white}$+0.128$ & \cellcolor{sky!34!white}$+0.280$ & \cellcolor{sky!28!white}$+0.166$ & \cellcolor{sky!20!white}$+0.077$ \\
\arrayrulecolor{mist}\cmidrule(lr){1-9}
\multirow{2}{*}{{\color{grp}\makecell[l]{fine-tuning\\point}}} & Chinese & \agree{0.917} & \cellcolor{sky!35!white}$+0.318$\,{\scriptsize\color{grp}/\,$0.291$} & \cellcolor{sky!32!white}$+0.233$ & \cellcolor{sky!34!white}$+0.287$ & \cellcolor{sky!40!white}$+0.518$ & \cellcolor{sky!38!white}$+0.420$ & \cellcolor{sky!32!white}$+0.245$ \\
 & English & \agree{0.917} & \cellcolor{sky!31!white}$+0.226$\,{\scriptsize\color{grp}/\,$0.211$} & \cellcolor{sky!27!white}$+0.146$ & \cellcolor{sky!26!white}$+0.130$ & \cellcolor{sky!31!white}$+0.218$ & \cellcolor{sky!29!white}$+0.175$ & \cellcolor{sky!27!white}$+0.143$ \\
\arrayrulecolor{mist}\cmidrule(lr){1-9}
\multirow{1}{*}{{\color{grp}\makecell[l]{precision}}} & transformers bf16 & \agree{0.878} & \cellcolor{sky!27!white}$+0.145$\,{\scriptsize\color{grp}/\,$0.156$} & \cellcolor{sky!20!white}$+0.080$ & \cellcolor{coral!46!white}\ownbox{$\mathbf{-0.309}$} & \cellcolor{coral!38!white}$-0.148$ & \cellcolor{coral!49!white}$\mathbf{-0.397}$ & \cellcolor{coral!52!white}$\mathbf{-0.509}$ \\
\arrayrulecolor{mist}\cmidrule(lr){1-9}
\multirow{2}{*}{{\color{grp}\makecell[l]{input}}} & band-limited & \agree{0.601} & \cellcolor{sky!48!white}$+1.145$\,{\scriptsize\color{grp}/\,$1.140$} & \cellcolor{sky!48!white}$+1.127$ & \cellcolor{sky!47!white}$+1.029$ & \cellcolor{coral!80!white}\ownbox{$\mathbf{-5.154}$} & \cellcolor{coral!77!white}$\mathbf{-4.130}$ & \cellcolor{coral!57!white}$-0.743$ \\
 & loudness-norm. & \agree{0.327} & \cellcolor{sky!60!white}$+3.65\phantom{0}$\,{\scriptsize\color{grp}/\,$3.66\phantom{0}$} & \cellcolor{sky!60!white}$+3.57\phantom{0}$ & \cellcolor{sky!60!white}$+3.63\phantom{0}$ & \cellcolor{sky!63!white}$+4.63\phantom{0}$ & \cellcolor{sky!62!white}$+4.43\phantom{0}$ & \cellcolor{sky!57!white}$+2.77\phantom{0}$ \\
\arrayrulecolor{ink}\midrule
\multicolumn{2}{l}{cross-entropy, reference codes} & --- & \multicolumn{1}{c}{$83.5$\,{\scriptsize\color{grp}/\,$83.4$}} & \multicolumn{1}{c}{$83.7$} & \multicolumn{1}{c}{$83.7$} & \multicolumn{1}{c}{$87.8$} & \multicolumn{1}{c}{$87.1$} & \multicolumn{1}{c}{$58.2$} \\
 
\bottomrule
\end{tabular}
\caption{Candidate encoders. Agreement: share of codes in codebooks 1--4 equal to the fp32 reference on the 512 test
utterances. $\Lambda$ (nats/frame, 16 codebooks) under calibration models trained on one label variant and under the
Talker. Cell color: {\color{coral!65!white}\rule{0.75em}{0.75em}} the model prefers the candidate to the fp32
reference ($\Lambda<0$), {\color{sky!60!white}\rule{0.75em}{0.75em}} it penalizes it ($\Lambda>0$); darker means larger
$|\Lambda|$ (log scale). The fp32 model was trained twice; the second seed (grey) shows the noise. Framed: a model scoring its own
labels. Bold: the preferences discussed in the text.}
\label{tab:lambda}
\end{table*}

\paragraph{Data.} Calibration models are trained on 107.6\,h of VCTK~\citep{vctk} (mic1) and AISHELL-3~\citep{aishell3}
from 263 speakers, cut into 10\,s chunks at 24\,kHz. Talker tests use 512 utterances (2--9\,s, 0.49\,h, 22{,}295 frames)
from 64 held-out speakers, 8 per speaker (speakers, not files, are held out; Appendix~\ref{app:candidates}). All confidence intervals are 95\% intervals from a speaker-level bootstrap ($B=1000$) that
shares resampling weights across models, so differences are paired.

\section{Is the label encoder Mimi's?}\label{sec:mimi}

\subsection{White-box evidence}\label{sec:whitebox}

The Qwen3-Omni checkpoint has no encoder or quantizer, but each generator has a learned embedding table per codebook
(2048 rows each). If the codes are Mimi's, each table should be predictable from Mimi's codebook vectors for the same
layer. We fit ridge regressions from Mimi's codebook $k$ to each Qwen table and report 5-fold cross-validated $R^2$
(Table~\ref{tab:whitebox}). Every Qwen table is predictable from the matching Mimi layer and from no other layer;
shuffling the rows destroys the fit. As a negative control, Qwen3-TTS-Tokenizer-12Hz~\citep{qwen3tts}, a codec with exactly the same
architecture (225 tensors with identical names and shapes, 223 of them with different values), predicts nothing.
This rules out a different tokenizer (iv). It cannot distinguish the same encoder (i) from a modified one (iii): a
fine-tuned encoder with a frozen quantizer would pass too.

\begin{table}[t]
\centering\small
\setlength{\tabcolsep}{3pt}
\renewcommand{\arraystretch}{1.15}
\begin{tabular}{@{}llcl@{}}
\toprule
\multicolumn{2}{@{}l}{embedding table \textcolor{grp}{\scriptsize codebooks}} & \tikz[baseline=0.2ex]{\draw[ink!55, line width=0.35pt] (0,0) -- (1.750,0);\draw[ink!55, line width=0.35pt] (0.187,0) -- (0.187,0.06);\node[font=\scriptsize, text=ink, inner sep=0.8pt, anchor=south] at (0.187,0.06) {0};\draw[ink!55, line width=0.35pt] (0.969,0) -- (0.969,0.06);\node[font=\scriptsize, text=ink, inner sep=0.8pt, anchor=south] at (0.969,0.06) {0.5};\draw[ink!55, line width=0.35pt] (1.750,0) -- (1.750,0.06);\node[font=\scriptsize, text=ink, inner sep=0.8pt, anchor=south] at (1.750,0.06) {1};}%
 & $R^2$ \\
\midrule
\multirow{3}{*}{{\color{grp}\makecell[l]{Qwen3-\\Omni}}} & Code2Wav \textcolor{grp}{\scriptsize 0--15} & \tikz[baseline=-0.6ex]{\fill[mist!45!white, rounded corners=1.2pt] (0,-0.075) rectangle (1.750,0.075);\draw[ink!55, line width=0.45pt] (0.187,-0.11) -- (0.187,0.11);\fill[navy, rounded corners=1.2pt] (0.578,-0.075) rectangle (1.203,0.075);} & 0.25--0.65 \\
 & Talker \textcolor{grp}{\scriptsize 0} & \tikz[baseline=-0.6ex]{\fill[mist!45!white, rounded corners=1.2pt] (0,-0.075) rectangle (1.750,0.075);\draw[ink!55, line width=0.45pt] (0.187,-0.11) -- (0.187,0.11);\fill[navy] (0.891,0) circle (0.072);} & 0.45 \\
 & MTP \textcolor{grp}{\scriptsize 1--15} & \tikz[baseline=-0.6ex]{\fill[mist!45!white, rounded corners=1.2pt] (0,-0.075) rectangle (1.750,0.075);\draw[ink!55, line width=0.45pt] (0.187,-0.11) -- (0.187,0.11);\fill[navy, rounded corners=1.2pt] (1.391,-0.075) rectangle (1.578,0.075);} & 0.77--0.89 \\
\arrayrulecolor{mist}\cmidrule(lr){1-4}
\multirow{2}{*}{{\color{grp}\makecell[l]{null}}} & rows shuffled & \tikz[baseline=-0.6ex]{\fill[mist!45!white, rounded corners=1.2pt] (0,-0.075) rectangle (1.750,0.075);\draw[ink!55, line width=0.45pt] (0.187,-0.11) -- (0.187,0.11);\draw[mist!25!ink, line width=0.8pt, -{Stealth[length=2.6pt]}] (0.141,0) -- (0.000,0);\fill[mist!25!ink] (0.141,0) circle (0.06);} & $\le -0.03$ \\
 & other layer & \tikz[baseline=-0.6ex]{\fill[mist!45!white, rounded corners=1.2pt] (0,-0.075) rectangle (1.750,0.075);\draw[ink!55, line width=0.45pt] (0.187,-0.11) -- (0.187,0.11);\draw[mist!25!ink, line width=0.8pt, -{Stealth[length=2.6pt]}] (0.141,0) -- (0.000,0);\fill[mist!25!ink] (0.141,0) circle (0.06);} & $\le -0.03$ \\
\arrayrulecolor{mist}\cmidrule(lr){1-4}
\multirow{1}{*}{{\color{grp}\makecell[l]{control}}} & Qwen3-TTS codec & \tikz[baseline=-0.6ex]{\fill[mist!45!white, rounded corners=1.2pt] (0,-0.075) rectangle (1.750,0.075);\draw[ink!55, line width=0.45pt] (0.187,-0.11) -- (0.187,0.11);\draw[coral!75!ink, line width=0.8pt, -{Stealth[length=2.6pt]}] (0.187,0) -- (0.000,0);\fill[coral!75!ink] (0.187,0) circle (0.06);} & $\le 0$ \\
\arrayrulecolor{ink} 
\bottomrule
\end{tabular}
\caption{White-box check: cross-validated $R^2$ of predicting each Qwen codebook embedding table from Mimi's codebook
vectors (bars: range over the table's codebooks). Nulls predict from shuffled rows or from another Mimi layer (maxima
over layers); the control from Qwen3-TTS-Tokenizer-12Hz, a codec with Mimi's architecture but other weights.}
\label{tab:whitebox}
\end{table}

\subsection{A decoder-based test is uninformative}\label{sec:decoder}

Code2Wav was trained on pairs $(E_Q(x), x)$ without text or speaker conditioning, so it is a natural witness. We swap
single codes of the fp32 reference codes for their second-nearest codeword and score each encoder by how well its own
quantization margins predict which swaps the decoder resists. Code2Wav ranks the fp32 reference above every candidate,
which looks like support for the same-encoder hypothesis. Calibration says otherwise: three Code2Wav-architecture
decoders trained from scratch, one of them only on random-5\% labels, all rank the reference first as well, and the
seed effect is about four times the label effect (Appendix~\ref{app:decoder}). The test cannot detect encoder changes
of the size of random 5\% or smaller, so we do not use it as evidence: decoders carry little information about the
encoder's quantization boundaries.

\subsection{Calibrated Talker likelihood}\label{sec:likelihood}

The Talker and MTP were trained with cross-entropy on $E_Q$'s codes, so label differences show up directly in their
likelihood. For each utterance we encode the audio with each candidate encoder $C$ and feed the codes to the Talker
(codebook 0) and the MTP (codebooks 1--15) under teacher forcing, with a plain read-aloud prompt
(``Please read the following sentence aloud:'' or its Chinese equivalent) and the preset voice Ethan; the voice
Chelsie gives the same conclusions. We report
\begin{equation}
\Lambda(C) = \frac{1}{T}\sum_{t}\sum_{k=0}^{15}\Big[\log p\big(c^{\mathrm{ref}}_{t,k}\big) - \log p\big(c^{C}_{t,k}\big)\Big]
\end{equation}
in nats per frame, where $c^{\mathrm{ref}}$ are the fp32 reference codes; $\Lambda(C)<0$ means the model prefers $C$'s
codes to the reference codes (implementation check in Appendix~\ref{app:candidates}).

Some codes may simply be easier to predict, which would bias $\Lambda$ regardless of the labels. We therefore train
small code language models with the Talker's factorization (a temporal Transformer for codebook 0, a depth Transformer
for codebooks 1--15; no text input), each on the 107.6\,h training set encoded by one encoder, and score them the same
way. A model trained on $C$'s labels should prefer $C$; two models trained on the same labels with different seeds
measure the noise.

Table~\ref{tab:lambda} shows that the test works where the decoder test did not: the seed effect is about 0.01, while
models trained on random-5\% or transformers bf16 labels prefer those labels by 0.27 and 0.31, about 40 times more.
Reading the Talker column against the calibration columns:

\begin{itemize}
\item \textbf{Half precision.} The Talker prefers transformers bf16 codes ($-0.509$, CI $[-0.654, -0.384]$). Every
model trained on fp32 labels penalizes these codes or prefers them far less (lower CI bound $\ge -0.238$, reached only
by the model trained on fully band-limited input); models trained on bf16 labels behave like the Talker.
\item \textbf{No loudness normalization:} every model, including the Talker, penalizes loudness-normalized codes heavily.
\item \textbf{Not along the fine-tuning directions we built:} the Talker penalizes the Chinese fine-tuning point
($+0.245$) as much as fp32 models do, while a model trained on its labels prefers them ($-0.234$, CI $[-0.31, -0.16]$).
\item \textbf{No dose response to random perturbations:} the Talker's penalties for the three random perturbations are
tiny ($+0.026$, $+0.023$, $+0.077$, at most 0.13\% of its cross-entropy). A model only prefers a perturbed encoder when
its labels lie exactly along that perturbation, so random probes cannot detect an unknown change.
\item \textbf{Band-limiting is mostly ruled out.} The Talker prefers band-limited codes on test speech
($-0.743$), but a check on speech the Talker generated itself shows that its labels contain less than 20\%
band-limited material (Appendix~\ref{app:robust}).
\end{itemize}

\section{Which implementation?}\label{sec:impl}

``Half precision'' is not one thing: libraries differ in kernel choices, casting points and the distance computation
in the quantizer, and in residual quantization one flipped code changes all later codebooks. We encoded the test set
with 22 implementation variants (Appendix~\ref{app:variants}) and trained further calibration models on their labels.

\begin{figure}[t]
\centering
\includegraphics[width=\columnwidth]{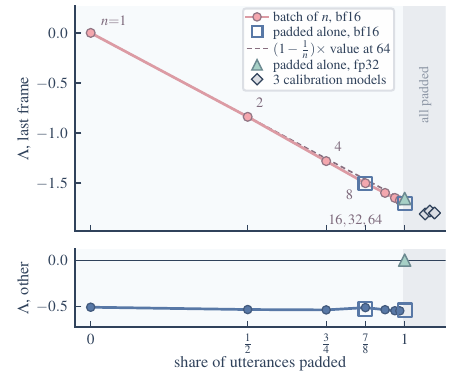}
\caption{The last frame carries the batch-size effect (Talker $\Lambda$, voice Ethan, nats/frame; both parts are averaged
over all frames). Top: in batches of $n$, a share $1-1/n$ of the utterances is padded, and the last-frame part follows
$(1-1/n)$ times its value at $n=64$ (dashed). Single utterances padded in bf16 (squares: as in a batch of 8, and to
twice their length) or in fp32 (triangle) land on the same line, and three calibration models trained on 10\,s chunks
without partial frames score the padded fp32 codes alike (diamonds). Bottom: the other frames stay at $-0.51$ to
$-0.55$ in bf16 and equal the reference in fp32.}
\label{fig:lastframe}
\end{figure}

\paragraph{The last frame carries a large, non-diagnostic effect.} Mimi's encoder is causal: appending zeros changes
nothing but the last frame (CPU fp32, maximum difference $4\times 10^{-7}$ elsewhere). transformers pads the last,
incomplete frame inside each convolution; when the input is padded, the deep layers instead see the network's actual
response to silence. Encoding in batches pads all but the longest utterance, and this is what made batched bf16 look
most Qwen-like in our first analysis. Figure~\ref{fig:lastframe} splits $\Lambda$ into the two parts: for batch size
$n$, the last-frame part equals the padded-last-frame effect times $1-1/n$ (predicted $-0.84$, $-1.26$, $-1.47$ for
$n=2,4,8$; measured $-0.84$, $-1.28$, $-1.50$), while the other frames stay at $-0.51$ to $-0.55$ for any batch size.
Every calibration model, including ones trained only on unpadded fp32 codes, prefers the padded last frame by about as
much as the Talker ($-1.81$, $-1.78$, $-1.80$ versus $-1.66$), because their 10\,s training chunks never end in a
partial frame. The last frame therefore says nothing about how Qwen encoded; from here on we exclude it from $\Lambda$.

\begin{figure*}[t]
\centering
\begin{minipage}[b]{0.705\textwidth}
\centering\footnotesize
\setlength{\tabcolsep}{2.7pt}
\renewcommand{\arraystretch}{1.12}
\begin{tabular}[b]{@{}ll rrrr @{\hspace{9pt}} r@{}}
\multicolumn{7}{@{}l}{\textbf{(a)}} \\[1pt]
\toprule
\multicolumn{2}{@{}l}{\color{grp}calibration model, labels from} & \multicolumn{3}{c}{transformers} & \code{moshi} & \\
\cmidrule(lr){3-5}\cmidrule(lr){6-6}
\multicolumn{2}{@{}l}{implementation} & fp32 & fp16 & bf16 & bf16 & \textbf{Talker} \\
\midrule
\input{figures/tab_impl_body.tex} 
\bottomrule
\end{tabular}
\end{minipage}\hfill
\begin{minipage}[b]{0.275\textwidth}
\raggedright{\footnotesize\textbf{(b)}}\\[1pt]
\includegraphics[width=\linewidth]{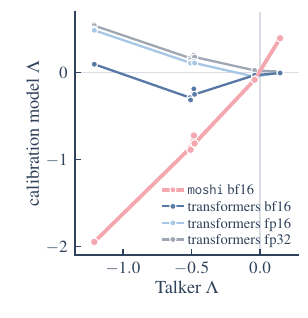}
\end{minipage}
\caption{Implementation fingerprint ($\Lambda$ excluding the last frame, nats/frame). (a) Encoder implementations
(Table~\ref{tab:impl}) in the order of the Talker's $\Lambda$ (voice Ethan; Chelsie differs by at most 0.02), scored by
four calibration models and by the Talker; cell color and frames as in Table~\ref{tab:lambda}. The Talker's column shares
its pattern only with the model trained on \code{moshi} bf16 labels. (b) The same values against the Talker's $\Lambda$:
only that model follows the Talker's order.}
\label{fig:impl}
\end{figure*}

\paragraph{The Talker ranks implementations like a model trained on \code{moshi} bf16 labels.} Figure~\ref{fig:impl}
shows that the Talker and a calibration model trained on \code{moshi} bf16 labels give the same order: \code{moshi} with
bf16 weights (streaming or not) far ahead, then the transformers bf16 variants and \code{moshi} with autocast, then
fp16, fp32 and TF32. The Talker's values are about 0.6 times the calibration model's; we only compare order and sign,
since the two models have different scales. No other calibration model shows this pattern: to models trained on
transformers bf16, fp16 or fp32 labels, \code{moshi} bf16 codes are harder to predict than the fp32 reference codes.
What matters is casting the weights to bf16: \code{moshi} with fp32 weights under autocast behaves like the
transformers variants. A code-level decomposition shows that the Talker's preference comes from the systematic part of
bf16 rounding: codes flipped in two runs of the same bf16 encoder with different batch sizes are preferred by 0.74 nats
each, codes flipped in only one run by about 0. On the training set, \code{moshi} bf16 codes agree with the fp32
reference on 82\% and with transformers bf16 codes on 85\% of codes in codebooks 1--4.

\paragraph{Robustness checks.} The result does not depend on the GPU, although the codes do: encoding on another GPU
changes 4--7\% of bf16 codes in codebooks 1--4, and the Talker prefers both sets of codes equally; over all 16
codebooks about a quarter of the codes change (Appendix~\ref{app:robust}). The fingerprint constrains the implementation but not the
weights: calibration models trained on bf16 labels from perturbed \code{moshi} weights keep the Talker's preference
profile, so it gives no bound on how far $E_Q$'s weights are from the public ones. As an independent check that
involves no decoding or re-encoding, the Talker's own sampled codes are fit best by the calibration model trained on
\code{moshi} bf16 labels (0.683 nats/frame above the model trained on fp32 labels, CI $[0.60, 0.75]$).
With \code{label\_check} (Section~\ref{sec:tools}) on FLEURS~\citep{fleurs}, the recipe is preferred over transformers
labels in all 10 of the Talker's speech languages, so the identification holds beyond Chinese and English.
Appendix~\ref{app:robust} gives the details.

\paragraph{Summary.} The white-box check rules out a different tokenizer (iv). Under hypothesis (i), which allows numerical implementation differences, $E_Q$ behaves like the
public Mimi encoder (first 16 codebooks) run with kyutai's \code{moshi} package and weights cast to bf16. Fixed
preprocessing (loudness normalization; band-limiting of 20\% or more of the data) and our two fine-tuning directions
are ruled out. Weight changes in other directions cannot be ruled out at any size: likelihood comparisons only say
which of two candidates the labels are closer to, and random probe directions are nearly orthogonal to an unknown
change. We identify behavior, not bit-exact weights.

\section{Fine-tuning on Mimi labels}\label{sec:finetune}

\subsection{The last-frame pitfall}\label{sec:lastframe}

\paragraph{transformers versus kyutai.} kyutai's reference encoder pads the input to a multiple of the frame size
(1920 samples) before the network, with a call to \code{pad\_for\_conv1d} in
\code{moshi/models/compression.py}; transformers' \code{MimiModel} lets each strided convolution pad its own edge, as in
EnCodec~\citep{encodec}. Both produce $\lceil n/1920\rceil$ frames. Zero-padding the input of each strided convolution
is not the same as padding the audio with silence: after the first layers silence is no longer zero, so the last frame
is encoded from different hidden states. On the 512 test utterances (CPU fp32, 32
codebooks) they agree on 99.99\% of tokens in earlier frames but on only 5.5\% in the last frame; the 6 utterances that
come out identical are those whose length is a multiple of 1920. The same mechanism makes batched encoding of
right-padded inputs disagree with single encoding~\citep{mimi_batch_43377}, since shorter items see real zeros after their end: with transformers
and 8 utterances per batch, earlier frames agree fully and last-frame tokens on 15.6\% (64 utterances). Padding the
input to a multiple of 1920 samples before encoding makes transformers match \code{moshi} on all frames (508 of 512
utterances identical) and makes batched and single encoding agree (100\%). We reported this to transformers and proposed this
change as the fix.\footnote{\url{https://github.com/huggingface/transformers/issues/50396}}

\begin{table*}[t]
\centering\small
\setlength{\tabcolsep}{2.55pt}
\renewcommand{\arraystretch}{1.12}
\begin{tabular}{@{}ll cccccc @{\hspace{6pt}} c@{}}
\toprule
 & & \multicolumn{3}{c}{without padding} & \multicolumn{3}{c}{tail-padded} & \\
\cmidrule(lr){3-5}\cmidrule(lr){6-8}
\multicolumn{2}{@{}l}{labels encoded with} & \thead{\footnotesize transformers\\\footnotesize fp32} &
\thead{\footnotesize transformers\\\footnotesize bf16} & {\footnotesize\code{moshi}} &
\thead{\footnotesize\textbf{\code{moshi} bf16}\\\footnotesize\textbf{(recipe)}} & \thead{\footnotesize transformers\\\footnotesize fp32} &
\thead{\footnotesize transformers\\\footnotesize bf16} & \thead{\footnotesize real\\\footnotesize recordings} \\
\midrule
\multirow{4}{*}{{\color{grp}\makecell[l]{Qwen3-\\Omni}}} & loud last frame (\%) & \loudtwo{49}{55} & \loudtwo{44}{53} & \loudtwo{0}{0.8} & \loudtwo{0}{0.3} & \loudtwo{0}{0} & \loudtwo{0.3}{0.3} & \loudtwo{0}{2.5} \\
 & last-frame jump (dB) & \cellcolor{coral!72!white}$+22.3$/$+7.0$ & \cellcolor{coral!70!white}$+21.4$/$+6.3$ & \cellcolor{sky!28!white}$-5.3$/$-5.3$ & \cellcolor{sky!30!white}$-5.7$/$-4.8$ & \cellcolor{sky!30!white}$-4.9$/$-5.7$ & \cellcolor{sky!30!white}$-5.6$/$-5.8$ & \cellcolor{sky!18!white}$-1.3$/$+0.7$ \\
 & \textcolor{grp}{CER / WER (\%)} & \textcolor{grp}{3.08/1.36} & \textcolor{grp}{3.19/1.24} & \textcolor{grp}{3.22/1.24} & \textcolor{grp}{3.11/1.16} & \textcolor{grp}{3.45/1.44} & \textcolor{grp}{3.45/1.24} & \textcolor{grp}{2.35/1.44} \\
 & \textcolor{grp}{similarity (sim.)} & \textcolor{grp}{.969/.958} & \textcolor{grp}{.968/.958} & \textcolor{grp}{.970/.959} & \textcolor{grp}{.969/.958} & \textcolor{grp}{.969/.958} & \textcolor{grp}{.968/.956} & \textcolor{grp}{---} \\
\arrayrulecolor{mist}\cmidrule(lr){1-9}
\multirow{3}{*}{{\color{grp}\makecell[l]{CSM-1B}}} & loud last frame (\%) & \loudone{46} & \loudone{50} & \loudone{0.6} & \textcolor{grp}{---} & \loudone{0} & \textcolor{grp}{---} & \textcolor{grp}{---} \\
 & last-frame jump (dB) & \cellcolor{coral!31!white}$+6.6$ & \cellcolor{coral!31!white}$+6.5$ & \cellcolor{sky!34!white}$-7.2$ & \textcolor{grp}{---} & \cellcolor{sky!33!white}$-6.9$ & \textcolor{grp}{---} & \textcolor{grp}{---} \\
 & \textcolor{grp}{WER (\%) / sim.} & \textcolor{grp}{2.9/.957} & \textcolor{grp}{2.7/.944} & \textcolor{grp}{2.0/.957} & \textcolor{grp}{---} & \textcolor{grp}{1.7/.953} & \textcolor{grp}{---} & \textcolor{grp}{1.4/---} \\
\arrayrulecolor{ink}
 
\bottomrule
\end{tabular}
\caption{Effect of the labels on LoRA fine-tuning. Qwen3-Omni: two voices, 3 training seeds, \code{moshi} in bf16,
Chinese/English; CSM-1B: Unsloth notebook, VCTK p239, 2 training seeds, \code{moshi} in fp32, English. Loud last
frame: share of generated sentences whose final 80\,ms is above $-20$\,dB relative to the loudest 80\,ms of the
sentence (bars from 0 to 60\%; upper bar Chinese);
last-frame jump: level of the final 80\,ms relative to the frame before (coral: a rise, blue: the normal decay).
Grey rows: intelligibility (CER/WER) and speaker similarity. No generation failed to stop.}
\label{tab:downstream}
\end{table*}

\paragraph{Effect on Qwen3-Omni fine-tuning.} We adapted the Talker with a LoRA~\citep{lora} to two new voices in one
run: AISHELL-3 speaker SSB0033 (Chinese, 23.4 minutes) in the slot of the preset voice Chelsie and VCTK speaker p239
(English, 23.5 minutes) in the slot of Aiden, with 3 training seeds per label set. Only the labels change between runs:
transformers fp32 without padding (the library default), transformers bf16 without padding, \code{moshi} bf16 without
padding, and our recipe (\code{moshi} bf16, tail-padded); a second set of runs adds transformers fp32 and bf16 with tail
padding. Unpadded transformers bf16 is what community training code does in practice~\citep{hert4,sca}. On 40
held-out sentences per voice we measure intelligibility (CER/WER), speaker similarity and the energy of the final
80\,ms relative to the loudest 80\,ms of the same sentence, with 80\,ms windows counted from the end of the audio; a
last frame above $-20$\,dB counts as loud. Appendix~\ref{app:finetune} gives the training and evaluation settings and
how the rates depend on this threshold.

\begin{figure}[t]
\centering
\includegraphics[width=\columnwidth]{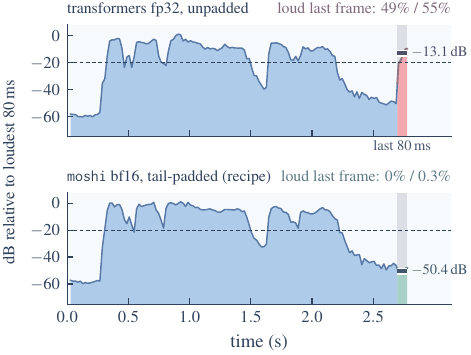}
\caption{Energy per 20\,ms of one held-out Chinese sentence generated after LoRA voice adaptation, same text and
sampling seed, with unpadded transformers fp32 labels (top) and with the recipe (bottom), in dB relative to the loudest
80\,ms of the sentence; the two utterances differ slightly because the models differ. The bar marks the level of the
shaded last 80\,ms, the dashed line $-20$\,dB. With the transformers labels the last 80\,ms jump by 36\,dB (from $-49$
to $-13$\,dB; Chinese average $+22$\,dB); with the recipe they fall to $-50$\,dB. Percentages: share of
generated sentences whose last 80\,ms is above $-20$\,dB (Chinese / English, mean of 3 training seeds;
Table~\ref{tab:downstream}).}
\label{fig:tail}
\end{figure}

For the four label sets of the first runs, intelligibility, speaker similarity, duration and stopping do not differ
(Table~\ref{tab:downstream}): all differences to the recipe have sentence-level bootstrap CIs containing 0 (CER
differences $-0.23$ to $+0.27$ points, similarity differences $\le 0.001$). For the two padded transformers label sets,
the CER, WER and similarity differences to the recipe also all have CIs containing 0. But with unpadded transformers
labels about half of the generated sentences end with an abnormal frame, a short noise burst: averaged over all
sentences, the last 80\,ms rise by 22\,dB in Chinese (from near silence, $-47$\,dB, to $-25$\,dB) and by 7\,dB in
English (from $-27$ to $-20$\,dB; Figure~\ref{fig:tail}). The burst is already in the labels: decoded with Code2Wav,
about half of the unpadded transformers fp32 training labels end above $-20$\,dB (Chinese 54\%, English 50\%), against
at most 1.5\% for \code{moshi} or padded labels and for the real training recordings; the fine-tuned Talker ends with a
burst about as often. With \code{moshi} labels the last frame decays normally, and so it does with
padded transformers labels in fp32 or bf16: padding alone removes the noise burst.

\paragraph{CSM-1B.} CSM-1B~\citep{csm} also uses Mimi codes, and its transformers training path encodes without
padding. It shows the same problem: the last frame of unpadded transformers labels costs the released model 119 nats
more than \code{moshi} labels, and after LoRA fine-tuning about half of the generations end with a loud last frame;
padding removes both (Table~\ref{tab:downstream}; details in Appendix~\ref{app:finetune}).

\subsection{Toolkit and the merged checkpoint}\label{sec:tools}

\paragraph{\code{label\_check} and \code{talker\_tune}.} \code{label\_check} scores a set of labels under the Talker
and reports $\Lambda$ against reference encodings, with and without the last frame; it shows at once whether labels were
padded and which implementation they resemble. \code{talker\_tune} prepares labels with the recipe, trains a LoRA (or
dense) adapter on the Talker and MTP, synthesizes speech, merges the adapter into a full checkpoint with the original layout, and evaluates CER/WER,
speaker similarity, end-of-utterance noise and stopping. In our runs with it (Appendix~\ref{app:finetune}), 2.5
minutes of speech per speaker already give the voice, and the voice is carried by the Talker LoRA, not the MTP. To
listen to or score 16-codebook Mimi codes, Code2Wav works better than Mimi's own decoder (Appendix~\ref{app:c2w}).

\paragraph{The merged checkpoint in chat.} Asked 20 questions per language, the full model merged from the 23-minute
LoRA answers text questions and spoken Chinese questions in the new voices (similarity 0.98) without noisy endings. Two
side effects showed up. First, the LoRA changes shared Talker weights, so the preset ids Chelsie and Aiden also speak
with the new voices. Second, for spoken English questions the English voice drifts toward the Chinese target speaker
(similarity to the voice's own text-question answers falls from 0.995 to 0.81 for p239): with one speaker per
language, the LoRA appears to tie the voice to the language of the prompt more than to the speaker id. Experiments
with more speakers (Appendix~\ref{app:drift}) lead to our recommendation for multilingual voices: train at least two
speakers of each gender per language, each for the full number of steps, or, where that is not possible, add
same-language spoken turns; the toolkit supports both.

\section{Related work}\label{sec:related}

Mimi~\citep{moshi} follows the SoundStream/EnCodec line of residual-quantized codecs~\citep{soundstream,encodec} with a
semantic first codebook; Qwen3-Omni~\citep{qwen3omni} uses its first 16 codebooks at 12.5\,Hz, where Qwen2.5-Omni used
a different tokenizer~\citep{qwen25omni}. Qwen3.5-Omni~\citep{qwen35omni} has no open weights at the time of writing.
Which encoder made the Talker's labels has been asked repeatedly in the Qwen3-Omni GitHub discussions from October 2025
to June 2026~\citep{qwen_d70,qwen_d149,qwen_d154,qwen_d176}, all without an official answer. The idea that the codec is Mimi came
from the community: in the Qwen2.5-Omni repository, abrar360 proposed it and SeungyounShin checked it by decoding
Qwen3-Omni codes with Mimi's decoder and by Talker/MTP losses~\citep{abrar_mimi,shin_forward_test}. Hert4 and the SCA
project published Talker training code that already uses Mimi to make labels~\citep{hert4,sca}, and Unsloth added
Talker and MTP fine-tuning kernels in 2026~\citep{unsloth_omni}, but no recipe for the labels. Recent full-duplex
systems built on Qwen3-Omni either keep the original Talker frozen~\citep{d2} or retrain it with labels from ``the Mimi
encoder''~\citep{duplexomni}. Our contribution
is to test how exact this identification is, which implementation matches, why the obvious decoder-based test fails,
and what goes wrong downstream with the default transformers encoder.

Even with the codec known, the codes of the same audio are not unique. \citet{codec_inconsistency} show that they change with the surrounding
context and with imperceptible perturbations, and that making them more consistent improves a codec language model; we
study how they change with numerical precision, the encoder implementation and the padding of the last frame. In
transformers, batched and single Mimi encoding were reported to give different codes~\citep{mimi_batch_35166,mimi_batch_43377},
without a comparison to kyutai's encoder or a link to the last frame. Undisclosed details of released models have been
inferred before, for example the training-data mixture of a BPE tokenizer from its merge list, checked on tokenizers
trained on known mixtures~\citep{bpe_mixture}; we identify the unreleased label encoder from the likelihoods the Talker
assigns, checked on models trained with known labels.

\section{Conclusion}\label{sec:conclusion}

Qwen3-Omni's Talker behaves as if its labels came from kyutai's Mimi encoder (first 16 codebooks) run with the \code{moshi} package and
bf16 weights; we identify this behavior, not bit-exact weights. A decoder-based test looks convincing but fails calibration,
so such a test should first be checked on models trained with known labels. transformers' \code{MimiModel}
agrees with kyutai's encoder on 99.99\% of the codes before the last frame, yet after LoRA fine-tuning on its unpadded
labels 44--55\% of Qwen3-Omni's sentences end in a noise burst, and CSM-1B shows the same problem;
padding the input before encoding removes them.

\section*{Limitations}

This is a behavioral identification. It cannot show that $E_Q(x)=\text{Mimi}(x)$ for every input or that the weights are
bit-identical; weight changes in directions we did not construct are undetectable by design. bf16 codes vary by
3--7\% in codebooks 1--4 across GPUs, PyTorch versions and tensor shapes, and by about 24\% over all 16 codebooks
across GPUs and batch sizes. Without Qwen's hardware, software and batching, which are unknown, its labels cannot be
reproduced code for code; in our tests this did not matter: the Talker prefers neither GPU's codes, and bf16 and fp32
labels gave the same LoRA results. The
downstream study covers LoRA voice adaptation with two speakers and about 47 minutes of speech (eight speakers for the
drift study); full fine-tuning, new languages and expressive styles are untested. A LoRA trained on read-aloud prompts
also changes the preset voices and, unless each language has more than one speaker of the same gender trained for the
full number of steps, lets an English voice fall back to a Chinese one when questions are spoken. The band-limiting
result only excludes fractions of 20\% or more.

\section*{Ethics Statement}

We use AISHELL-3 (Apache-2.0), VCTK 0.92 (CC BY 4.0) and FLEURS (CC BY 4.0), the Mimi weights (CC BY 4.0), Qwen3-Omni
(Apache-2.0) and CSM-1B (Apache-2.0) under their licenses, and the evaluation models (Paraformer, Parakeet-TDT, Whisper,
WavLM-SV) and the Qwen3-TTS-Tokenizer-12Hz control under their own. The toolkit we release adapts a speech model to a new voice
from a few minutes of recordings, so it could be used to imitate a person without their consent. The eight sample voices
we release are AISHELL-3 and VCTK speakers and follow the licenses of these corpora; we ask users to adapt the model to
another person's voice only with that person's consent.

\bibliography{refs}

\appendix

\section{Candidate encoders and scoring details}\label{app:candidates}

\paragraph{Candidate encoders.} \emph{Random perturbations} add Gaussian noise to the weights, scaled by each tensor's
norm, with the scale found by bisection so that 2\%, 5\% and 10\% of codes in codebooks 1--4 change. \emph{Fine-tuning
points} interpolate toward a real fine-tuning direction (25 Adam steps on Chinese or English speech with the quantizer
frozen) to 8\% disagreement. Band-limiting resamples the input from 24\,kHz to 16\,kHz and back to 24\,kHz; loudness
normalization scales it to $-23$\,LUFS.

\paragraph{Held-out speakers.} 170 AISHELL-3 \code{test/} speakers also appear in \code{train/}, so we hold out
speakers, not files.

\paragraph{Talker scoring.} Our implementation follows SeungyounShin's \code{forward\_test.py}~\citep{shin_forward_test}
and the code paths of the transformers class \code{Qwen3OmniMoe}\allowbreak\code{ForConditionalGeneration}; greedy
generation followed by teacher-forced rescoring reproduces the argmax for 97--100\% of codebook-0 frames.

\section{Encoder variants and implementation results}\label{app:variants}\label{app:impl}

Table~\ref{tab:impl} describes every label variant in this report and where its results are; the $\Lambda$ values of
the implementation variants are printed in Figure~\ref{fig:impl}. All variants were encoded with PyTorch 2.9.1 and
transformers 4.57.6. With the same PyTorch on an H100, \code{moshi} 0.2.13 reproduces our H100 recipe codes bit for bit
(512 of 512 utterances).

\begin{table*}[t]
\centering\small
\setlength{\tabcolsep}{3.5pt}
\renewcommand{\arraystretch}{1.08}
\begin{tabular}{@{}llll@{}}
\toprule
\multicolumn{2}{@{}l}{label variant} & how it is produced & \textcolor{grp}{results} \\
\midrule
\multirow{8}{*}{\color{grp}Figure~\ref{fig:impl}} & \code{moshi} bf16 & \code{loaders.get\_mimi}, weights \code{.to(torch.bfloat16)} & \textcolor{grp}{} \\
 & \code{moshi} bf16, streaming & same, one frame per step & \textcolor{grp}{} \\
 & transformers bf16, eager & \code{MimiModel} in bf16, \code{eager} attention & \textcolor{grp}{} \\
 & transformers bf16, sdpa & \code{MimiModel} in bf16, \code{sdpa} attention & \textcolor{grp}{also Table~\ref{tab:lambda}} \\
 & transformers + autocast & \code{MimiModel}, fp32 weights under bf16 autocast & \textcolor{grp}{} \\
 & \code{moshi} + autocast & \code{loaders.get\_mimi}, fp32 weights under bf16 autocast & \textcolor{grp}{} \\
 & transformers fp16 & \code{MimiModel} in fp16 & \textcolor{grp}{} \\
 & transformers TF32 & \code{MimiModel} in fp32, TF32 matrix multiplies & \textcolor{grp}{} \\
\arrayrulecolor{mist}\cmidrule(lr){1-4}
\multirow{7}{*}{\color{grp}elsewhere} & fp32 reference & \code{MimiModel} in fp32; all $\Lambda$ are relative to its codes & \textcolor{grp}{Section~\ref{sec:setup}} \\
 & \code{moshi} fp32 & \code{loaders.get\_mimi} in fp32; as the reference except the last frame & \textcolor{grp}{Section~\ref{sec:lastframe}} \\
 & transformers bf16, batch $n$ & $n$ utterances per batch, right-padded ($n=2,\dots,64$) & \textcolor{grp}{Figure~\ref{fig:lastframe}} \\
 & \code{moshi} bf16, perturbed & weights plus noise (2\% or 10\% of codes change), then bf16 & \textcolor{grp}{Table~\ref{tab:decay}} \\
 & transformers, padded & fp32 or bf16, input zero-padded at the tail & \textcolor{grp}{Table~\ref{tab:downstream}} \\
 & recipe & \code{moshi} bf16, input zero-padded at the tail & \textcolor{grp}{Figure~\ref{fig:recipe}, Table~\ref{tab:downstream}} \\
 & candidate encoders & random, fine-tuning, band-limited, loudness-normalized & \textcolor{grp}{Section~\ref{sec:setup}, Table~\ref{tab:lambda}} \\
\arrayrulecolor{ink}
 
\bottomrule
\end{tabular}
\caption{Label variants and how each is produced, grouped by where their results are. All encode 24\,kHz audio, keep the
first 16 codebooks and, unless stated otherwise, encode each utterance on its own without padding.}
\label{tab:impl}
\end{table*}

\section{Decoder test details}\label{app:decoder}

For each code in the fp32 reference codes we swap in the second-nearest codeword and ask whether the decoder's output
moves away from the original audio (log-mel L1 over the next 3 frames). A decoder trained on an encoder's labels should
prefer that encoder's choices most where the choice is close to a quantization boundary, so we score each reference
encoder $R$ by the AUC with which $R$'s own quantization margins predict the decoder's preferences (codebooks 1--4;
codebook 0 and codebooks $\ge 9$ carry no signal in a power pilot). Code2Wav gives the fp32 reference the highest AUC
(0.701), above every candidate (e.g.\ $\Delta A=-0.0104$, CI $[-0.0129, -0.0080]$ for random 5\%), and a sweep
along each perturbation direction peaks near the reference.

The reference, however, sits at the center of the candidate set, so a decoder could rank it first regardless of its
labels. We therefore trained three Code2Wav-architecture decoders from scratch for 30{,}000 steps: two on fp32
reference labels with different seeds, and one on random-5\% labels with the same seed as the first. All three rank
the reference first, including the one that never saw a reference label. The seed effect is about four times the
label effect: for $\Delta A$ of random 5\% it is $+0.0077$ versus $+0.0019$, and for the peak position along the
random-5\% direction $+1.13$ versus $+0.32$. Fine-tuning a copy of Code2Wav on random-5\% labels for 4{,}000 steps did
not move its peak either. A symmetric statistic that avoids anchoring on the reference's probes has no power at this
scale.

\section{Robustness checks}\label{app:robust}

\paragraph{Band-limiting.} On test speech the Talker's $\Lambda=-0.743$ for band-limited input looked like partial
band-limited training data. On speech the Talker generated itself (190 sentences decoded by Code2Wav, i.e.\ in its own
training distribution), the Talker penalizes band-limited input ($+0.923$) more than an fp32 model does ($+0.393$),
while a model trained with only 20\% band-limited chunks already prefers it ($-0.651$, CI $[-0.83, -0.47]$). So the
Talker's labels contain less than 20\% band-limited material; smaller fractions are untested, and the $-0.743$ on test
speech remains unexplained.

\paragraph{GPU and PyTorch version.} Encoding the same variants on an H100 instead of an RTX PRO 6000 changes 4--7\% of bf16 codes in
codebooks 1--4 (fp32 codes are virtually identical), similar to changing tensor shapes on one GPU. The Talker prefers
the H100 codes just as much ($\Lambda$ of \code{moshi} bf16 codes: $-1.232$ versus $-1.212$), and the calibration model
trained on PRO 6000 codes prefers both equally ($-1.978$ versus $-1.955$). The part of bf16 rounding that varies across
GPUs is not what the Talker learned, so labels can be reproduced on any GPU, though not code for code: re-encoding the
38{,}741 training chunks on an H100 instead of the PRO 6000 (same \code{moshi} and PyTorch versions, batch size 16
instead of 64) keeps 99.0\% of the codebook-0 codes, 92.8\% in codebooks 1--4 and 76.0\% over all 16 codebooks; only
38\% of frames agree in all 16 codebooks. GPU and batch size are not separated in this comparison. Changing PyTorch
from 2.9.1 to 2.15 on the H100 changes a similar share of the recipe codes (6.5\% in codebooks 1--4, 0.9\% in codebook
0); we did not score these codes with the Talker.

\paragraph{Perturbed weights.} Can the \code{moshi} bf16 preference bound how far $E_Q$'s weights are from the public
ones? We perturbed \code{moshi}'s encoder weights to 2\% and 10\% fp32 code disagreement, encoded in bf16 and trained a
calibration model on each. Before running, we fixed the reading: if perturbed-label models no longer prefer
\code{moshi} bf16 codes, the fingerprint is weight-specific and gives a bound. Table~\ref{tab:decay} shows the other
outcome: the preference for \code{moshi} bf16 codes moves by $+7\%$/$-6\%$ and the profile of ratios by at most 0.05.
The profile is set by the implementation and is insensitive to weight changes of this size, so it gives no bound.
Pairwise comparisons do not help either: the Talker prefers unperturbed to perturbed \code{moshi} bf16 codes, but so
does every model, including those trained on the perturbed labels, because perturbed labels are harder to model in
general.

\begin{table*}[t]
\centering\small
\setlength{\tabcolsep}{3.4pt}
\renewcommand{\arraystretch}{1.12}
\begin{tabular}{@{}ll r@{\hspace{3pt}}l cccc@{}}
\toprule
\multicolumn{2}{@{}l}{scored by} & \multicolumn{2}{c}{\thead{$\Lambda$ of \code{moshi} bf16 codes\\{\scriptsize\color{grp}95\% CI}}} &
\thead{transformers\\bf16} & \thead{\code{moshi}\\autocast} & \thead{transformers\\fp16} & \thead{transformers\\TF32} \\
\midrule
\multirow{3}{*}{{\color{grp}\makecell[l]{calibration\\models}}} & trained on \code{moshi} bf16 labels & \cellcolor{coral!68!white}$-1.955$ & {\scriptsize\color{grp}$[-2.35, -1.61]$} & \cellcolor{coral!58!white}$0.45$ & \cellcolor{coral!54!white}$0.42$ & \cellcolor{coral!12!white}$0.04$ & \cellcolor{sky!30!white}$-0.20$ \\
 & \quad same, weights perturbed 2\% & \cellcolor{coral!69!white}$-2.088$ & {\scriptsize\color{grp}$[-2.53, -1.73]$} & \cellcolor{coral!58!white}$0.45$ & \cellcolor{coral!54!white}$0.42$ & \cellcolor{coral!11!white}$0.03$ & \cellcolor{sky!31!white}$-0.21$ \\
 & \quad same, weights perturbed 10\% & \cellcolor{coral!68!white}$-1.832$ & {\scriptsize\color{grp}$[-2.28, -1.48]$} & \cellcolor{coral!52!white}$0.40$ & \cellcolor{coral!49!white}$0.37$ & \cellcolor{coral!9!white}$0.01$ & \cellcolor{sky!36!white}$-0.25$ \\
\arrayrulecolor{mist}\cmidrule(lr){1-8}
{\color{grp}Talker} & voice Ethan & \cellcolor{coral!63!white}$-1.213$ & {\scriptsize\color{grp}$[-1.42, -1.01]$} & \cellcolor{coral!54!white}$0.42$ & \cellcolor{coral!51!white}$0.39$ & \cellcolor{coral!11!white}$0.03$ & \cellcolor{sky!21!white}$-0.12$ \\
\arrayrulecolor{ink}
 
\bottomrule
\end{tabular}
\caption{Fingerprint decay: calibration models trained on bf16 labels from \code{moshi}, with and without perturbed
encoder weights, keep the Talker's preference profile ($\Lambda$ in nats/frame, last frame excluded). The four right
columns give $\Lambda(c)/\Lambda(\text{\code{moshi} bf16})$ for codes $c$ from the named implementation: coral when the
sign is the same, blue when not; other colors as in Table~\ref{tab:lambda}.}
\label{tab:decay}
\end{table*}

\paragraph{The Talker's own samples.} We scored the Talker's raw sampled codes (190 sentences) under each calibration
model. All models fit their own labels on test speech almost equally well (log-likelihood $-82.09$ to $-82.27$
nats/frame, except the band-limited one), so differences reflect which labels the samples resemble. Relative to the
model trained on fp32 labels, the model trained on \code{moshi} bf16 labels is higher by 0.683 (CI $[0.60, 0.75]$),
the one trained on transformers bf16 labels by 0.235 and a second fp32 seed by 0.205, while the models trained on
Chinese fine-tuning point and band-limited labels are lower ($-0.618$, $-6.20$). On real \code{moshi} bf16 codes the
corresponding gaps are 0.715 and 0.443, almost the same as the 0.683 and 0.448 seen on the samples.

\paragraph{Other languages.} With \code{label\_check} on FLEURS~\citep{fleurs} (40 test utterances in each of the
Talker's 10 speech languages, 16\,kHz audio resampled, base model, preset voice Ethan), the recipe beats transformers
fp32 labels by 0.8--4.9 nats/frame and transformers bf16 labels by 0.4--4.0 on all frames but the last (all CIs
exclude 0), and cannot be told from unpadded \code{moshi} bf16 labels ($|\Delta|\le 0.09$, all CIs contain 0).
Padding lowers the last-frame NLL of transformers labels by 15--50 nats in every language.

\section{Fine-tuning setup and toolkit ablations}\label{app:finetune}

\paragraph{Qwen3-Omni voice adaptation.} SSB0033 has 458 and p239 has 462 training utterances, with 40 held-out
utterances each. The Thinker is frozen (the prompt uses only its token embeddings). A rank-16 LoRA on the Talker's
attention and shared expert and the MTP's attention and MLP (6.3M parameters) is trained for 800 steps with batch 8 and
learning rate $10^{-4}$ (cosine) on the 16-codebook cross-entropy plus end-of-speech. The runs with tail-padded
transformers labels use the same settings. Hert4's script~\citep{hert4} and the SCA training code~\citep{sca} both
encode with transformers' \code{MimiModel} in the Talker's dtype (bf16), without padding. We sample each held-out
sentence with 3 seeds (top-$k$ 50, temperature 0.9), decode with Code2Wav, and measure CER with
Paraformer~\citep{paraformer} (Chinese), WER with Parakeet-TDT~\citep{tdt} (English), speaker similarity to the
held-out recordings with WavLM-SV~\citep{wavlm}, and the energy of the final 80\,ms in dB relative to the loudest
80\,ms window of the sentence (windows counted from the end of the audio; above $-20$\,dB counts as a loud last frame).
Absolute rates depend on this definition. We repeated the count at thresholds of $-30$, $-25$, $-20$ and $-15$\,dB with
windows of 40, 80 and 160\,ms. \code{moshi} labels (padded or not) give at most 2\% loud last frames in every setting
in Chinese, and in English at $-20$ and $-15$\,dB. The held-out recordings of p239, however, end at a higher relative
level ($-26$\,dB on average in the last 80\,ms; SSB0033 $-44$\,dB), so lower thresholds count normal English endings as
loud: with the recipe, 11\% at $-25$\,dB and 49\% at $-30$\,dB with 80\,ms windows, 90\% at $-30$\,dB with 160\,ms
windows. In English, that \code{moshi} labels give few loud last frames therefore holds only at $-20$ and $-15$\,dB. The
difference between unpadded transformers fp32 labels and the recipe is positive in all 24 combinations of threshold,
window and language (bootstrap over seeds and sentences; every CI excludes 0).
For reference, the base model with
preset voices has CER 12.8\% and WER 8.5\%.

\paragraph{CSM-1B.} CSM uses all 32 Mimi codebooks, and transformers' CSM training path encodes each audio segment with
\code{codec\_model.encode} without padding, which is what the Unsloth~\citep{unsloth} fine-tuning notebook uses. Under the released
model (160 VCTK utterances), the last frame of transformers labels costs 199.4 nats summed over the 32 codebooks
against 80.1 with \code{moshi} fp32 labels ($+119$, CI $[+109, +130]$); padding the input brings it to 80.2
(difference $+0.08$, CI $[-0.23, +0.42]$). Other frames are unaffected ($-0.003$ nats). Unlike the Qwen Talker, CSM
prefers fp32 labels: transformers bf16 labels cost $+0.39$ nats/frame (CI $[+0.21, +0.56]$) on the other frames.
Precision should therefore be checked per model by likelihood, not assumed. Fine-tuning shows the same symptom as the
Talker (LoRA rank 32 on the backbone and depth decoder, 400 steps, batch 8, learning rate $2\times10^{-4}$; 40 held-out
sentences sampled twice): about half of the generations end with a loud last frame with unpadded transformers labels
and almost none with padded or \code{moshi} labels (fp32 unpadded minus \code{moshi}: $+46$ points, CI $[+38, +53]$).
WER and speaker similarity do not separate the label sets at this sample size.

\paragraph{\code{talker\_tune}.} The Thinker is loaded without its decoder layers, since read-aloud training only needs
its embeddings: the Talker alone takes 7.2\,GB, LoRA training peaks at 14\,GB allocated on transformers 5.20 and runs
under a 22\,GB per-process cap (0.45\,s per step), so a 24\,GB card suffices. Table~\ref{tab:toolkit} repeats the
voice adaptation with the toolkit (one seed, one sample per held-out sentence; its evaluator uses Paraformer for
Chinese and Whisper-large-v3-turbo~\citep{whisper} for English). A new speaker id and a replaced preset voice work equally well; 2.5
minutes of speech per speaker already give the voice, and more data mainly lowers the error rates; the Talker LoRA
carries the voice, while an MTP-only LoRA leaves English similarity at 0.88 and brings back noisy endings.

\begin{table}[t]
\centering\footnotesize
\setlength{\tabcolsep}{2.8pt}
\renewcommand{\arraystretch}{1.12}
\begin{tabular}{@{}ll cccc@{}}
\toprule
\multicolumn{2}{@{}l}{run} & \thead{CER\\zh} & \thead{WER\\en} & \thead{similarity\\zh / en} &
\thead{loud last\\frame zh / en} \\
\midrule
\multirow{2}{*}{{\color{grp}\makecell[l]{voice}}} & new ids & \cellcolor{coral!8!white}3.3 & \cellcolor{coral!11!white}3.7 & \cellcolor{coral!8!white}.969/.960 & \loudtwo{0}{0} \\
 & preset slots & \cellcolor{coral!12!white}3.8 & \cellcolor{coral!6!white}3.0 & \cellcolor{coral!6!white}.970/.963 & \loudtwo{0}{0} \\
\arrayrulecolor{mist}\cmidrule(lr){1-6}
\multirow{3}{*}{{\color{grp}\makecell[l]{data}}} & 10 min & \cellcolor{coral!19!white}4.8 & \cellcolor{coral!27!white}5.8 & \cellcolor{coral!9!white}.965/.959 & \loudtwo{0}{0} \\
 & 5 min & \cellcolor{coral!27!white}5.8 & \cellcolor{coral!28!white}5.9 & \cellcolor{coral!10!white}.966/.957 & \loudtwo{0}{2.5} \\
 & 2.5 min & \cellcolor{coral!26!white}5.7 & \cellcolor{coral!28!white}5.9 & \cellcolor{coral!12!white}.968/.955 & \loudtwo{0}{5} \\
\arrayrulecolor{mist}\cmidrule(lr){1-6}
\multirow{2}{*}{{\color{grp}\makecell[l]{LoRA}}} & Talker only & \cellcolor{coral!6!white}3.0 & \cellcolor{coral!16!white}4.3 & \cellcolor{coral!10!white}.966/.958 & \loudtwo{0}{0} \\
 & MTP only & \cellcolor{coral!68!white}11.3 & \cellcolor{coral!11!white}3.7 & \cellcolor{coral!68!white}.962/.880 & \loudtwo{18}{10} \\
\arrayrulecolor{ink}\midrule
\multicolumn{2}{@{}l}{\color{grp}real recordings} & \textcolor{grp}{2.5} & \textcolor{grp}{3.4} & \textcolor{grp}{---} & \loudtwo{0}{2.5} \\
 
\bottomrule
\end{tabular}
\caption{\code{talker\_tune} runs (\%; 23 minutes of speech per speaker unless noted; 800 steps, one seed, each of 40
held-out sentences per speaker sampled once).
Cell color: worse than the best run (coral, darker is worse); bars as in Table~\ref{tab:downstream}. Real recordings are
scored with the same ASR. Differences of about one point are within noise.}
\label{tab:toolkit}
\end{table}

\section{Voice drift in chat}\label{app:drift}

\paragraph{The merged checkpoint in chat.} We asked the full model, merged from the 23-minute LoRA, 20 single-turn
questions per language with transformers; the Thinker answers and the Talker speaks (about 50\,s per answer). For text
questions in both languages and spoken Chinese questions the answers have similarity 0.98 to the target speaker and no
noisy endings. The preset ids Chelsie and Aiden also speak with the new voices (similarity to the target 0.98 and
0.97). For spoken English questions (synthesized by the base model in a preset voice) the voice drifts: against the
centroid of each voice's own text-question answers, similarity falls from 0.995 to 0.81 for p239 and to 0.74 for
Aiden. The base model, asked the same spoken questions, keeps Aiden's voice (0.994 vs 0.993 for text questions), so
the drift comes from the LoRA. The drifted English answers are in the voice of the Chinese target speaker (similarity
0.97 to SSB0033, which WavLM-SV barely separates from Chelsie, the voice the new speaker rows start from). Spoken user
turns paired at random across languages did not fix this: with 200 such turns (training sentences read by the base
model in two preset voices, Thinker hidden states precomputed) in half of the samples, English answers switched between
the two voices in every condition, including read-aloud (similarity 0.88 vs 0.96), while Chinese was unaffected.
vLLM-Omni~\citep{vllm_omni} 0.30 serves the merged checkpoint with no change, taking \code{speaker} in the chat request (10 text
questions per language: similarity 0.985 and 0.988 for the new voices; the preset ids again take the new voices).

\begin{figure}[t]
\centering
\includegraphics[width=\columnwidth]{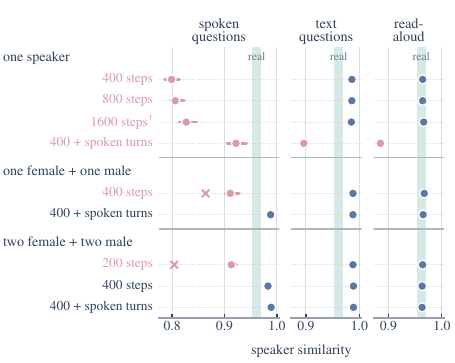}
\caption{English voices after LoRA training with different speaker sets per language (training steps per speaker;
spoken turns: same-language spoken questions in half of the samples), one panel per kind of prompt. Dots: speaker
similarity to the speaker's own held-out recordings, mean of three seeds (bars: seed range); coral: below the range of
real recordings (green band), i.e.\ the voice has drifted. Crosses: the female voice p239. $^\dagger$Overfits (CER
22\%, WER 15\%). The Chinese voices reach 0.95--0.99 in every run.}
\label{fig:drift}
\end{figure}

\paragraph{More than one speaker per gender and language.} To test the one-speaker explanation we retrained with four
speakers per language (AISHELL-3: SSB0033 and SSB0435 female, SSB0603 and SSB0590 male; VCTK: p239 and p236 female,
p259 and p286 male; at most 23.5 minutes each; female ids start from Chelsie, male ids from Ethan), three seeds per run
(the first row of Figure~\ref{fig:drift} repeats the 23-minute LoRA). Each speaker answered the 20 questions of its
language as text and as speech and read 20 held-out sentences. Since the LoRA only touches the Talker, the Thinker
answers and the hidden states of the spoken questions were computed once and shared by all runs; we synthesized the
first 64 answer tokens (about 20\,s), 120 answers per batch with left-padded prompts (in fp32, batched and single
greedy decoding give identical codes). Similarity is to the centroid of the speaker's own held-out recordings (real
recordings, leave-one-out: 0.953--0.970); English WER is from Whisper-large-v3-turbo. Figure~\ref{fig:drift} shows the
results. With one speaker per language the voice drifts in every seed (59 of 60 spoken English answers are nearest to
SSB0033), and two or four times as many steps do not help; four times as many overfit (Chinese CER 19--22\%, English
WER 15--21\%). Two female and two male speakers per language, at the same 400 steps per speaker, remove the drift when
trained on read-aloud prompts only (0.983 vs 0.989 for text questions), with error rates unchanged. The same speakers
at 200 steps each do not: both English female voices drift to a Chinese female voice (p239: 0.804), the male voices do
not. With one female and one male speaker per language only the female voice drifts (p239: 0.864); spoken turns of the
same language in half of the samples fix it (0.988), whereas with one speaker per language the same turns make English
text answers and read-aloud drift as well (0.896 and 0.885). Female and male ids start from different presets, so
gender and initialization cannot be told apart here. These results give the recommendation in
Section~\ref{sec:tools}; the toolkit supports both options (initial voice per speaker, same-language spoken turns). One
speaker recorded in both languages remains untested.

\section{Code2Wav as a decoder for Mimi codes}\label{app:c2w}

Each transposed convolution in Code2Wav (transformers implementation) crops $k-s$ samples from both ends, so the output
is 555 samples (23\,ms) shorter than 1920 per frame, a deficit already reported for streamed
decoding~\citep{c2w555_sglang,c2w555_mstar}; the left crop also makes the output lead its input by the same 555
samples. Shift it back before comparing with reference audio, or spectral and intelligibility metrics come out too low.
Aligned, Code2Wav decodes the recipe codes (16 codebooks) of the 512 test utterances about as well as Mimi's decoder
does from all 32 codebooks, and better than Mimi's decoder from the same 16 (Table~\ref{tab:c2w}). On codes sampled
from the base Talker the two decoders give the same voice (similarity between the decodings 0.97/0.98). To listen to or
score 16-codebook Mimi codes, we therefore use Code2Wav (objective metrics only; no listening test). This is a
statement about the decoder, not evidence about the encoder (Section~\ref{sec:decoder}).

\begin{table}[t]
\centering\small
\setlength{\tabcolsep}{3.1pt}
\renewcommand{\arraystretch}{1.12}
\begin{tabular}{@{}lcccc@{}}
\toprule
decoder & \thead{log-mel\\L1 (dB)} & similarity & \thead{CER/WER\\(\%)} & STOI \\
\midrule
Code2Wav, 16 cb. & \cellcolor{coral!6!white}3.2/3.0 & \cellcolor{coral!6!white}.981/.982 & \cellcolor{coral!11!white}3.3/3.0 & \cellcolor{coral!23!white}.94/.86 \\
Mimi, 16 cb. & \cellcolor{coral!58!white}4.5/4.0 & \cellcolor{coral!58!white}.965/.966 & \cellcolor{coral!58!white}3.9/3.9 & \cellcolor{coral!58!white}.92/.83 \\
Mimi, 32 cb. & \cellcolor{coral!37!white}3.9/3.6 & \cellcolor{coral!19!white}.977/.980 & \cellcolor{coral!6!white}3.3/2.9 & \cellcolor{coral!6!white}.95/.86 \\
\midrule
\textcolor{grp}{original audio} & \textcolor{grp}{---} & \textcolor{grp}{---} & \textcolor{grp}{3.1/2.5} & \textcolor{grp}{---} \\
 
\bottomrule
\end{tabular}
\caption{Decoding the recipe codes of the 512 test utterances (Chinese/English; Code2Wav's output shifted back by 555
samples). Similarity is to the original audio; STOI: short-time objective intelligibility~\citep{stoi}. Cell color: worse than the best decoder (coral, darker is worse).
``cb.'': codebooks.}
\label{tab:c2w}
\end{table}

\end{document}